\documentclass[particles,review,submit,moreauthors]{Definitions/mdpi}

\usepackage{bm}
\usepackage{float}
\renewcommand{\dagger}{\mathchar"2279\relax}
\newcolumntype{Y}{>{\raggedright\arraybackslash}X}

\makeatletter
\fancypagestyle{plain}{%
  \fancyhf{}%
  \fancyhead{\null\vspace{8pt}}%
  \fancyfoot[R]{}%
}
\makeatother

\firstpage{1}
\pubvolume{1}
\issuenum{1}
\articlenumber{0}
\pubyear{2026}
\copyrightyear{2026}
\datereceived{ }
\daterevised{ }
\dateaccepted{ }
\datepublished{ }

\Title{Neutrino Anomalies and Tensions: Experimental Status and Theoretical Interpretations}
\Author{Doris Bar\v{c}ot $^{1,}$*, Raoul Serao $^{2,}$* and Antonio Capolupo $^{3}$}
\AuthorNames{Doris Bar\v{c}ot, Raoul Serao and Antonio Capolupo}
\address{%
$^{1}$ \quad Division of Experimental Physics, Ru{\dj}er Bo\v{s}kovi\'{c} Institute, Bijeni\v{c}ka cesta 54, 10000 Zagreb, Croatia; \texttt{dbarcot@irb.hr}\\
$^{2}$ \quad Department of Physics, Faculty of Science, University of Zagreb, Bijeni\v{c}ka cesta 32, 10000 Zagreb, Croatia; \texttt{rserao@phy.hr}\\
$^{3}$ \quad Dipartimento di Fisica E.~R. Caianiello, Universit\`a di Salerno, and INFN -- Gruppo Collegato di Salerno, Via Giovanni Paolo II 132, 84084 Fisciano (SA), Italy; \texttt{capolupo@sa.infn.it}}
\corres{Correspondence: dbarcot@irb.hr (D.B.); rserao@phy.hr (R.S.)}

\abstract{Neutrino oscillations successfully describe flavor conversion across a wide range of experiments, yet the origin of several excesses and deficits remains unresolved. This review connects these observations and long-baseline parameter tensions to their proposed physical explanations, clarifying what remains open and how it can be tested. We examine reactor and gallium measurements, accelerator appearance and disappearance searches, and atmospheric constraints, focusing on sterile neutrinos, non-standard interactions, and decoherence in open-system dynamics. For each mechanism, we assess what it modifies, which observations it can address, and which theoretical assumptions and independent constraints limit its viability. Recent results have restricted several interpretations and weakened some earlier tensions; no single extension provides an uncontested account of all observations. We identify the flavor, baseline, energy, and event-topology signatures needed to distinguish competing explanations and assess whether the remaining anomalies require new physics or reflect unresolved flux, interaction, and detector uncertainties.}


\begin{document}

\section{Introduction}
\label{sec:intro}

Neutrino oscillations provide direct evidence that neutrinos have nonzero masses and that lepton flavor is not conserved during propagation. Their discovery established one of the first experimental results requiring an extension of the original Standard Model of particle physics. The theoretical description is based on the mixing of neutrino states with definite flavor and definite mass, first developed by Pontecorvo and subsequently generalized by Maki, Nakagawa, and Sakata \cite{Pontecorvo1957,Maki1962}. The observation of atmospheric neutrino oscillations by Super-Kamiokande, the resolution of the solar neutrino problem by the Sudbury Neutrino Observatory (SNO), and the confirmation of reactor antineutrino disappearance by KamLAND established the experimental foundations of this framework \cite{SuperK1998,SNO2001,KamLAND2003}.

In the standard three-neutrino description, the flavor states $\nu_e$, $\nu_\mu$, and $\nu_\tau$, corresponding to electron, muon, and tau neutrinos, are coherent superpositions of three mass eigenstates $\nu_1$, $\nu_2$, and $\nu_3$. Their mixing is described by the unitary Pontecorvo--Maki--Nakagawa--Sakata (PMNS) matrix. In vacuum, flavor-transition probabilities depend on the neutrino energy $E$, the propagation distance $L$, the mixing parameters, and the mass-squared differences $\Delta m_{ij}^{2}=m_i^{2}-m_j^{2}$, where $m_i$ and $m_j$ are neutrino masses. The observed oscillation pattern is governed by two independent mass-squared splittings, associated with solar and atmospheric neutrino oscillations. Matter interactions modify this evolution and play an essential role in the interpretation of solar and long-baseline accelerator measurements.

The three-neutrino framework successfully describes a broad range of observations covering different energies, propagation distances, production mechanisms, and detection techniques. Global analyses provide increasingly precise determinations of the mixing angles and mass-squared differences. Important questions nevertheless remain open, including the neutrino mass ordering, which specifies the relative arrangement of the three masses, and the value of the leptonic charge--parity (CP) phase $\delta_{\rm CP}$, which controls CP-violating effects in the standard mixing matrix \cite{PDG2024,NuFit2024}. These open questions do not undermine the established oscillation picture. They instead define the next targets of precision neutrino physics.

Alongside these successes, several experiments have reported event excesses, rate deficits, or differences between preferred oscillation parameters. These observations are commonly grouped under the term neutrino anomalies. They do not, however, constitute a single experimental problem. Some concern the appearance of events associated with a flavor different from that produced at the source. Others involve the disappearance of an expected neutrino flux. Further discrepancies emerge when measurements obtained with different experimental configurations are analyzed within a common oscillation model. Their interpretation depends on the accuracy of neutrino flux predictions, interaction cross sections, detector responses, background estimates, and statistical procedures. A discrepancy between observed and predicted event rates is therefore not sufficient evidence for new neutrino oscillations.

The most extensively investigated appearance anomalies originate from the Liquid Scintillator Neutrino Detector (LSND) and MiniBooNE experiments. LSND reported an excess of electron-antineutrino candidates in a predominantly muon-antineutrino beam. MiniBooNE subsequently observed an excess of electron-like events at short baselines. These results motivated the hypothesis of an additional oscillation frequency associated with a mass-squared splitting much larger than those measured in solar and atmospheric experiments. A minimal realization is the $3+1$ model, which introduces a fourth neutrino mass eigenstate with predominantly sterile character. A sterile neutrino does not participate in standard weak interactions but can affect flavor transitions through mixing with the active states.

This interpretation faces substantial experimental constraints. In a consistent oscillation model, the mixing parameters responsible for short-baseline appearance also affect neutrino disappearance. Searches involving electron and muon neutrinos therefore provide complementary tests of the same hypothesis. Their combined results place the simplest sterile-neutrino explanation under significant tension. The MicroBooNE experiment has also used liquid-argon detector technology to distinguish electron-induced events from important background contributions. At the 95\% confidence level, its two-beam analysis excludes the displayed LSND 99\% allowed region and most of the displayed MiniBooNE 95\% allowed region within the tested single-light-sterile-neutrino model \cite{MicroBooNE2025}. These constraints do not exclude every possible origin of the observed excesses. They show that an appearance signal cannot be interpreted independently of disappearance measurements and detector-specific backgrounds.

A second class of anomalies concerns reactor electron antineutrinos. The reactor antineutrino anomaly emerged from a deficit in measured event rates relative to revised theoretical flux predictions. This discrepancy motivated searches for additional electron-antineutrino disappearance at short distances. Its interpretation is particularly sensitive to the predicted antineutrino spectra produced by nuclear fission. Different calculations of these spectra lead to different expected event rates and therefore to different assessments of the anomaly. Recent measurements have improved the determination of reactor antineutrino yields and exposed limitations of widely used flux models. In particular, Daya Bay has reported a total flux compatible with some updated predictions, while significant spectral discrepancies remain \cite{DayaBay:2025ngb}. Reactor data consequently test both neutrino oscillation hypotheses and the reliability of nuclear calculations. Comparisons based on energy-dependent spectral ratios provide additional information because they reduce the dependence on absolute flux normalization.

The gallium anomaly provides a different experimental challenge. Radiochemical source experiments, including GALLEX and SAGE, measured fewer electron-neutrino capture events than expected. The Baksan Experiment on Sterile Transitions (BEST) confirmed a substantial deficit using an intense radioactive source and two detection zones. These results have frequently been interpreted in terms of electron-neutrino disappearance into a sterile state. However, the predicted capture rates depend on the nuclear transition strengths and on the theoretical treatment of neutrino absorption by gallium. Recent calculations have revisited these ingredients and examined whether changes in the capture cross section could reduce the discrepancy without introducing additional neutrino states \cite{GalliumFactorization2026}. The gallium results therefore require a joint assessment of nuclear physics uncertainties, source measurements, and independent oscillation constraints. Their statistical significance within a given cross-section model must be distinguished from the significance of any particular new-physics interpretation.

Long-baseline (LBL) accelerator experiments address a related but distinct issue. Experiments such as T2K and NOvA measure neutrino appearance and disappearance over distances of hundreds of kilometers. Their results constrain the atmospheric mass-squared splitting, the mixing angle governing muon-neutrino disappearance, the neutrino mass ordering, and the CP-violating phase. Different experimental configurations can produce different parameter preferences because of their energy spectra, matter effects, detector responses, and correlations between oscillation parameters. The first joint T2K--NOvA analysis has improved the combined parameter determination while showing that the interpretation of CP violation remains correlated with the assumed mass ordering \cite{T2KNOvA2025}. Differences between individual best-fit values should not automatically be classified as anomalies. Their physical relevance depends on a quantitative compatibility analysis that includes statistical fluctuations and correlated systematic uncertainties.

These experimental results motivate several extensions of standard neutrino phenomenology. Light sterile neutrinos introduce additional propagating states and oscillation frequencies. Non-standard neutrino interactions (NSI) modify neutrino production, detection, or propagation through interactions beyond the standard weak processes. In matter, propagation NSI can alter the effective flavor-evolution Hamiltonian and change the oscillation probabilities measured by long-baseline experiments. Non-unitary effective mixing provides another possible description when the observed active-neutrino sector is part of a larger theory. Neutrino decay and interactions with additional light particles can also modify observable event distributions. These mechanisms have different implications for their dependence on energy, propagation distance, matter density, and experimental channel. Their predictions must be compared with all relevant constraints rather than with a single anomalous dataset.

An additional possibility arises when neutrino propagation is described as an open quantum system. In the standard treatment, neutrino flavor evolution is coherent and unitary. Interactions with unobserved degrees of freedom, or suitable effective descriptions of unresolved dynamics, may instead produce decoherence. This can suppress interference between propagation components and modify the flavor-transition probabilities. Lindblad-type evolution equations provide a framework for describing Markovian open-system dynamics while imposing conditions such as trace preservation and complete positivity. Decoherence models can generate effects that differ from those produced by additional mass eigenstates or modified matter interactions. However, an extra damping term is not, by itself, evidence of an environmental mechanism. Its physical origin, energy dependence, and consistency with existing oscillation measurements must be established.

Quantum field theory and quantum-information methods offer further tools for examining the assumptions underlying neutrino mixing and coherence. Field-theoretical descriptions of flavor mixing address the relation between flavor and mass representations and the structure of the associated quantum states. Quantum-information approaches characterize coherence, correlations, and their evolution during propagation. These methods can clarify the physical content of effective oscillation and decoherence models, although they do not automatically imply observable corrections to standard oscillation probabilities \cite{SeraoJStat2025,SeraoCurved2023,SeraoTorsion2024,Serao:2026vxh,Capolupo:2024ckb,Capolupo:2021cnt}. Any proposed correction must be connected to a measurable quantity and assessed against the precision of current experiments.

The central difficulty is not the absence of theoretical explanations for individual anomalies. It is the limited ability of many proposed mechanisms to describe different datasets simultaneously. A model that reproduces an appearance excess may conflict with disappearance bounds. An explanation based on modified propagation must remain consistent with solar, atmospheric, reactor, and accelerator measurements. A mechanism that changes event production must reproduce the observed energy and final-state distributions without violating independent constraints. These requirements provide a common basis for comparing otherwise different theoretical proposals.

This review examines the principal neutrino anomalies within that framework. Section~\ref{sec:anomalies} presents the experimental observations, emphasizing the measured quantities, the assumptions entering their interpretation, and the constraints from complementary datasets. Section~\ref{sec:interpretations} discusses the main theoretical mechanisms, with particular attention to light sterile neutrinos, non-standard interactions, and open-system decoherence. Other scenarios are considered where they provide distinct observable signatures. Section~\ref{sec:outlook} identifies measurements capable of distinguishing competing explanations and discusses the prospects for resolving the remaining discrepancies. The objective is not to introduce a new global fit, but to establish which interpretations remain compatible with the available evidence and which experimental signatures can discriminate among them.

\section{Experimental Anomalies and Tensions}
\label{sec:anomalies}
An absolute-rate deficit, a spectral distortion, and an electron-like excess probe different observables. Charged-current (CC) interactions identify neutrino flavor through the accompanying charged lepton. Neutral-current (NC) interactions do not identify that flavor. Electron-like showers can also originate from photons. These distinctions determine which new-physics interpretation an experiment can test.

\begin{figure}[tb]
\centering
\includegraphics[width=0.96\textwidth]{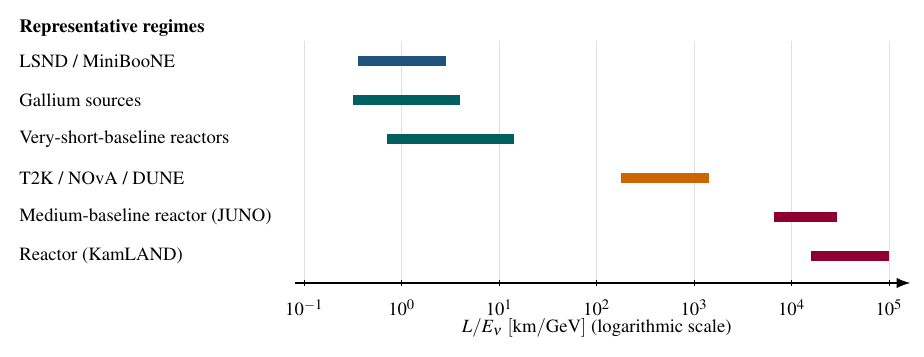}
\caption{Representative ranges of the baseline-to-energy ratio $L/E_\nu$, with $L$ the propagation distance and $E_\nu$ the neutrino energy. The axis is logarithmic, and $1\,\mathrm{m/MeV}=1\,\mathrm{km/GeV}$. Bands show approximate kinematic coverage, not confidence intervals or detector acceptances. Source and reactor measurements have different energies and geometries even when their ratios overlap. Atmospheric data span a much wider range; at TeV energies, Earth matter effects also matter. DUNE denotes the planned beam configuration.}
\label{fig:LE}
\end{figure}

Throughout, $L$ is the source-to-detector distance and $E_\nu$ is the neutrino energy. The states $\nu_\alpha$ and $\bar\nu_\alpha$ denote neutrinos and antineutrinos of flavor $\alpha=e,\mu,\tau$, corresponding to electron, muon, and tau flavor. A mass eigenstate $\nu_i$ has mass $m_i$, and $\Delta m^2_{ij}\equiv m_i^2-m_j^2$ is the mass-squared splitting of states $i$ and $j$. Figure~\ref{fig:LE} shows representative values of $L/E_\nu$. Quoted significances use Gaussian-equivalent standard deviations, denoted by $\sigma$; confidence levels and Bayesian credible intervals are distinguished in the text.

\subsection{Reactor antineutrinos}
\label{sec:reactor}
Reactors produce electron antineutrinos through beta decays of fission products. At MeV energies, the main detection channel is inverse beta decay (IBD),
\begin{equation}
\bar\nu_e+p\longrightarrow e^++n,
\label{eq:ibd}
\end{equation}
where $p$ and $n$ denote a proton and neutron, and $e^+$ denotes a positron. The rate depends on the emitted spectrum, the IBD cross section, and the detector response. The original reactor antineutrino anomaly was a roughly $6\%$ SBL rate deficit relative to the Huber--Mueller predictions \cite{Mention2011,Huber2011,Mueller2011}. These predictions mainly convert measured aggregate beta spectra into antineutrino spectra. The $^{238}$U contribution instead uses a summation of individual beta branches. Summation calculations depend on fission yields, decay data, and their correlated uncertainties.

A separate discrepancy is the ``5~MeV bump.'' Daya Bay, RENO, Double Chooz, and NEOS observed an excess around $4$--$6$~MeV in reconstructed prompt energy \cite{DayaBay:2015lja,RENO:2015ksa,DoubleChooz:2014kuw,NEOS2017}. The prompt visible energy $E_{\rm prompt}$ includes positron kinetic energy and annihilation photons. To leading order in IBD kinematics, $E_{\rm prompt}\simeq E_\nu-0.78$~MeV. The offset follows from the neutron--proton and electron masses. The bump and the integrated deficit need not have the same origin. A shared spectral feature without baseline dependence can arise from flux modeling.

Fuel evolution helps test that possibility. Daya Bay measured how the yield and spectrum change with the $^{239}$Pu fission fraction. Its 2023 analysis favored the tested summation model over Huber--Mueller for yield evolution, while spectral discrepancies remained \cite{DayaBay:2022jnn}. Its 2025 measurement used $4.7\times10^6$ IBD candidates and compared total and isotopic yields with several models \cite{DayaBay:2025ngb}. Agreement in total yield did not imply agreement in spectral shape. Isotope-dependent flux uncertainties must therefore be included with their correlations.

The rate significance also depends on the flux covariance. Using the summation calculation of Periss\'e et al. \cite{Perisse2023}, a 2026 reanalysis found a $2.2\sigma$ discrepancy; other updated flux models gave smaller discrepancies \cite{giunti2026}. The same study compared reactor and gallium results under specified error models. Enlarging the assumed gallium uncertainty weakens their tension, but does not identify its physical source.

Relative spectral measurements test baseline dependence with less reliance on absolute flux predictions. STEREO compared detector segments near a predominantly $^{235}$U source. PROSPECT-I compared spectra at different baselines at the High Flux Isotope Reactor. Neither final analysis established sterile-neutrino oscillations \cite{STEREO2023,PROSPECT2025}. DANSS moves one detector between positions below a reactor; its 2025 report extends earlier exclusions \cite{DANSS:2018fnn,DANSS2025}. NEOS compares with an externally measured reference spectrum, which requires corrections for fuel composition and detector response \cite{NEOS2017}.

These tests retain uncertainties from relative calibration, geometry, backgrounds, and spectral covariance. They lose sensitivity when the new phase is too small or oscillations average similarly at all positions. Reactor data exclude large regions of eV-scale mixing, while leaving flux-model discrepancies and some sterile-neutrino parameter space open \cite{ReactorReview2024,PROSPECT2025}.

\subsection{The gallium anomaly}
\label{sec:gallium}
GALLEX and SAGE calibrated their detectors with $^{51}$Cr and $^{37}$Ar electron-capture sources. The emitted electron neutrinos produce radioactive germanium through
\begin{equation}
\nu_e+{}^{71}\mathrm{Ga}\longrightarrow{}^{71}\mathrm{Ge}+e^-,
\label{eq:gacapture}
\end{equation}
where $e^-$ is an electron and $^{71}$Ga and $^{71}$Ge are the target and daughter nuclei. The extracted germanium atoms are counted radiochemically. The measured production rates were below predictions based on source activities and capture cross sections \cite{GALLEX:1994rym,GALLEX1998,SAGE1999,SAGE:2005ysf,Kaether:2010ag,Giunti:2010zu}. Extraction and counting efficiencies enter this comparison; no event-by-event neutrino spectrum is reconstructed.

BEST used two concentric gallium zones around a $^{51}$Cr source. With its reference cross section, the measured-to-predicted rate ratios were $R_{\rm in}=0.79\pm0.05$ and $R_{\rm out}=0.77\pm0.05$ for the inner and outer zones \cite{BEST2022}. Both show deficits. Their relative rates show no significant spatial variation. Correlated source and cross-section uncertainties affect both zones. A normalization shift or unresolved oscillations can produce similar deficits after averaging over the source and target volumes.

Conventional capture calculations infer the dominant ground-state matrix element from the $^{71}$Ge electron-capture half-life. Excited states at 175 and 500~keV add less certain contributions \cite{Haxton1998,Kostensalo2019}. Omitting them reduces the prediction but leaves a deficit. A 2026 calculation with self-consistent electron wave functions found combined source-data significances of $5.5\sigma$ and $4.7\sigma$ for two excited-state prescriptions, and $3.8\sigma$ for the ground state alone \cite{Cadeddu2025}. These values depend on the capture calculation and covariance model.

A study published in September 2026 relaxed the factorization of leptonic and nuclear wave functions \cite{GalliumFactorization2026}. It used phenomenologically constrained Gamow--Teller transition densities that reproduce the measured half-life. The resulting lower capture cross sections can accommodate the source rates. The half-life fixes one weighted nuclear integral; it does not fix every capture amplitude when the spatial dependence is treated explicitly. Independent nuclear calculations and cross-section tests are needed to validate this proposal.

The deficit therefore persists under conventional capture calculations, while its nuclear interpretation remains open. With those conventional rates, the minimal $3+1$ explanation conflicts with reactor spectral ratios, solar data, and parts of the KATRIN and MicroBooNE exclusions. Changing the cross section changes the inferred mixing region. Proposed source measurements using neutrino--electron scattering would test the emitted flux without the same gallium nuclear matrix elements \cite{ciuffoli2025}.

\subsection{Short-baseline accelerator appearance}
\label{sec:sbl_accelerator}
\label{sec:sbl_appearance}
LSND searched for $\bar\nu_\mu\to\bar\nu_e$ appearance using positive-muon ($\mu^+$) decay at rest. It identified IBD through a prompt positron and delayed neutron capture. At a baseline near 30~m and energies of tens of MeV, LSND reported $87.9\pm22.4$ (statistical) $\pm6.0$ (systematic) excess events, or about $3.8\sigma$ \cite{LSND2001}. The averaged appearance probability was $(0.264\pm0.067\pm0.045)\%$, with uncertainties in the same order. At $L/E_\nu\sim1$~m/MeV, the measured solar and atmospheric splittings cannot generate this signal. KARMEN used the same source and detection channels at a shorter baseline. Its 15 candidates agreed with $15.8\pm0.5$ expected background events, excluding part of LSND's oscillation region \cite{KARMEN:2002zcm}.

MiniBooNE tested appearance at 541~m with pion- and kaon-decay beams and energies of several hundred MeV. Its final combined neutrino and antineutrino samples contained $638.0\pm52.1$ (statistical) $\pm122.2$ (systematic) excess electron-like events, quoted at $4.8\sigma$ \cite{MiniBooNE2021}. Electrons and converted photons can produce similar Cherenkov signatures. The excess therefore depends on particle identification and interaction modeling. LSND's positron--neutron coincidence and MiniBooNE's electromagnetic showers are distinct observables.

MicroBooNE uses a liquid-argon time projection chamber to resolve interaction vertices, shower starts, and associated hadrons. Its first electron-neutrino searches did not confirm the tested MiniBooNE-inspired excess models \cite{MicroBooNE2022a,MicroBooNE2022b}. Its 2025 analysis combined the Booster Neutrino Beam (BNB) and the Neutrinos at the Main Injector (NuMI) beam. Their different electron-neutrino fractions reduce appearance--disappearance cancellations. In the tested single-light-sterile model, its 95\% exclusion contour covers the displayed LSND 99\% allowed region and most of the displayed MiniBooNE 95\% allowed region \cite{MicroBooNE2025}. The analysis uses the modified frequentist $CL_s$ measure, which accounts for limited exclusion sensitivity; the contour corresponds to $CL_s=0.05$. The result constrains that oscillation model without determining the physical origin of either excess.

MicroBooNE's inclusive single-photon study, published in May 2026, found overall agreement with its prediction, with goodness-of-fit $p$-value $p=0.11$ \cite{MicroBooNESinglePhoton2026}. Here a $p$-value is the probability, under the tested hypothesis, of obtaining a test statistic at least as incompatible as observed. A sample without visible protons showed $93\pm22$ (statistical) $\pm35$ (systematic) events above prediction, concentrated below 600~MeV shower energy. Its local significance was just above $2\sigma$. This feature does not establish a photon explanation of MiniBooNE. Such a comparison requires the different targets, acceptances, and event selections.

JSNS$^2$ tests the LSND channel with a pulsed decay-at-rest source and gadolinium-loaded scintillator. Its first result used 2022 data and was published in 2026. It found two events for $2.3\pm0.4$ expected background events; the LSND-motivated signal expectation was only $1.1\pm0.5$ events \cite{JSNS2First2026}. The first sample is too small for a decisive test. Further exposure can provide an independent comparison in the same production and detection channels.

\subsection{Disappearance searches and consistency}
\label{sec:sbl_numu_disappearance}
\label{sec:ap-disap-tension}
An oscillation explanation of appearance also predicts depletion of the parent flavor. Joint MiniBooNE/SciBooNE analyses at two BNB baselines found no significant $\nu_\mu$ or $\bar\nu_\mu$ disappearance \cite{MiniBooNESciBooNE2012nu,MiniBooNESciBooNE2012antinu}. MINOS/MINOS+ combined CC and NC samples at 1.04 and 735~km and found no additional oscillation component \cite{MINOS2019}. Both detectors contribute to their eV-scale sensitivity. Oscillations at the near detector and correlated uncertainties must be included.

T2K far-detector and NOvA dual-baseline sterile searches add complementary constraints \cite{T2K:2019efw,NOvA:2024imi}. The first ICARUS BNB search, released as a 2026 preprint using 2022--2023 data, found no significant $\nu_\mu$ disappearance at 600~m \cite{ICARUS:2026lvh}. Its 90\% confidence contours use an effective two-flavor approximation. Flux and interaction uncertainties limit the result. A joint analysis with the Short-Baseline Near Detector (SBND) at 110~m is intended to constrain those uncertainties.

In minimal $3+1$ mixing, appearance requires both electron- and muon-flavor components in the additional mass eigenstate. Electron- and muon-disappearance searches constrain those same components. Global analyses show strong tension between the required appearance signal and disappearance limits \cite{Dentler:2018sju,Acero2024}. Its significance depends on the selected data, likelihoods, and model. Section~\ref{sec:sterile} derives the amplitude relation responsible for this tension.

\subsection{Atmospheric constraints}
\label{sec:atmospherics}
Atmospheric neutrinos span broad energies and Earth path lengths. For a mostly sterile fourth state $\nu_4$ with $\Delta m^2_{41}>0$, Earth's NC matter potential can resonantly enhance $\bar\nu_\mu$ conversion at TeV energies. Energy, density, and zenith angle enter separately. Vacuum $L/E_\nu$ scaling is then insufficient. IceCube separates neutrino and antineutrino contributions statistically, without identifying their charges event by event.

IceCube's 10.7-year analysis of 0.5--100~TeV events found a best fit at $\sin^2(2\theta_{24})=0.16$ and $\Delta m^2_{41}=3.5$~eV$^2$, with a null-hypothesis $p$-value of $3.1\%$ \cite{IceCube2024}. The angle $\theta_{24}$ controls muon--sterile mixing in the convention of Eq.~\eqref{eq:Vparam}. This is a mild preference within that fit. Its strength depends on the other mixing parameters, Earth matter profile, and flux uncertainties. The $p$-value is neither an exclusion confidence level nor the probability that the sterile model is true.

\subsection{Long-baseline accelerator comparisons}
\label{sec:LBL_acc}
T2K and NOvA measure $\nu_\mu$ disappearance and $\nu_e$ appearance in beams enriched in neutrinos or antineutrinos. Appearance probes the Dirac CP phase $\delta_{\rm CP}$ and the mass ordering. Disappearance mainly constrains the atmospheric mixing angle $\theta_{23}$ and splitting $\Delta m^2_{31}$. The angle $\theta_{13}$ controls electron-flavor mixing with the third mass eigenstate and is independently constrained by reactors. Normal ordering has $m_3>m_2>m_1$; inverted ordering has $m_2>m_1>m_3$. The two experiments have different baselines, energies, and matter effects, so their compatibility requires a joint fit.

Their first joint analysis reports Bayesian credible intervals and ordering Bayes factors \cite{T2KNOvA2025}. A Bayes factor is the ratio of the marginal likelihoods for two hypotheses. With the default reactor constraint on $\theta_{13}$, the factor for inverted over normal ordering is 1.3. Adding a reactor constraint involving the atmospheric splitting weakly favors normal ordering instead. Neither comparison establishes the ordering. Under inverted ordering, the reported $3\sigma$-equivalent credible interval excludes the CP-conserving phases $\delta_{\rm CP}=0,\pi$; under normal ordering, a broad range remains allowed. This conditional Bayesian interval is not a frequentist discovery significance. The joint data are compatible with three-flavor oscillations.

JUNO's September 2026 preprint combines its reactor spectrum with external LBL constraints on the atmospheric splitting \cite{JUNO2026}. It favors normal ordering with a reported one-sided significance of at least $2.1\sigma$ across $\delta_{\rm CP}$, and separately quotes a Bayes factor of 10.18. Both results depend on the combined datasets and analysis assumptions. New-physics studies can test the robustness of such inferences, but an ordering preference alone does not identify a new propagation mechanism.

\section{Theoretical Interpretations}
\label{sec:interpretations}
A neutrino measurement combines production, propagation, and detection. Additional states, new interactions, and open-system dynamics modify different parts of this chain. A viable model must use the same parameters across all relevant channels. In the equations below, an asterisk denotes complex conjugation and a dagger denotes Hermitian conjugation.

\subsection{Light sterile neutrinos}
\label{sec:sterile}
A sterile flavor has no Standard Model weak interactions. A mass eigenstate with an active admixture can still be produced and detected. In the $3+1$ model, the fourth mass eigenstate $\nu_4$ is mostly sterile. SBL applications usually place its splitting $\Delta m^2_{41}$ near the eV$^2$ scale \cite{Abazajian2012,Acero2024}. Flavor and mass states obey
\begin{equation}
|\nu_\alpha\rangle=\sum_{i=1}^4 V_{\alpha i}^*|\nu_i\rangle,
\qquad \alpha=e,\mu,\tau,s,
\label{eq:flavor}
\end{equation}
where $s$ labels the sterile flavor and $V$ is the unitary four-state mixing matrix. We use the parametrization
\begin{align}
V&=U_{34}(\theta_{34},\delta_{34})U_{24}(\theta_{24},\delta_{24})U_{14}(\theta_{14},0)V_{3\nu},\nonumber\\
V_{3\nu}&=U_{23}(\theta_{23},0)U_{13}(\theta_{13},\delta_{\rm CP})U_{12}(\theta_{12},0).
\label{eq:Vparam}
\end{align}
Each $U_{ij}$ is a unitary rotation in the $(i,j)$ plane. Its $2\times2$ block has diagonal entries $\cos\theta_{ij}$ and off-diagonal entries $\sin\theta_{ij}e^{-i\delta_{ij}}$ and $-\sin\theta_{ij}e^{i\delta_{ij}}$; all remaining directions are unchanged. The angles $\theta_{12}$, $\theta_{13}$, and $\theta_{23}$ are the standard solar, reactor, and atmospheric mixing angles. The angles $\theta_{14}$, $\theta_{24}$, and $\theta_{34}$ describe the additional rotations. The phases $\delta_{24}$ and $\delta_{34}$ supplement $\delta_{\rm CP}$. The matrix $V_{3\nu}$ contains the unitary three-flavor matrix $U$ as its upper $3\times3$ block and a unit entry in the sterile direction. Majorana phases cancel from oscillation probabilities and are omitted. In the three-active-state models below, $|\nu_\alpha\rangle=\sum_{i=1}^3 U_{\alpha i}^*|\nu_i\rangle$. The sterile convention gives $|V_{e4}|^2=\sin^2\theta_{14}$ and $|V_{\mu4}|^2=\cos^2\theta_{14}\sin^2\theta_{24}$.

The SBL approximation neglects the solar and atmospheric phases:
\begin{equation}
\left|\frac{\Delta m^2_{21}L}{4E_\nu}\right|\ll1,
\qquad
\left|\frac{\Delta m^2_{31}L}{4E_\nu}\right|\ll1.
\label{eq:sbl_limit}
\end{equation}
The masses and splittings remain nonzero. At $L/E_\nu\sim1$~km/GeV, the atmospheric phase is about $3\times10^{-3}$. The approximation is useful when these phases and matter corrections are below the required precision. LBL and atmospheric fits generally require both effects.

In this vacuum limit, unitarity gives the amplitude for $\nu_\alpha\to\nu_\beta$, up to an unobservable common phase:
\begin{equation}
\mathcal A_{\alpha\beta}=\delta_{\alpha\beta}
+V_{\beta4}V_{\alpha4}^*\left(e^{-2i\Delta_{41}}-1\right),
\qquad
\Delta_{41}\equiv\frac{\Delta m^2_{41}L}{4E_\nu}.
\label{eq:sbl_amp}
\end{equation}
Here $\delta_{\alpha\beta}$ is the Kronecker delta: one for equal indices and zero otherwise. The dimensionless quantity $\Delta_{41}$ is the additional oscillation phase. The probabilities $P_{\alpha\beta}\equiv|\mathcal A_{\alpha\beta}|^2$ are
\begin{align}
P_{ee}&=1-4|V_{e4}|^2(1-|V_{e4}|^2)\sin^2\Delta_{41},\nonumber\\
P_{\mu\mu}&=1-4|V_{\mu4}|^2(1-|V_{\mu4}|^2)\sin^2\Delta_{41},\nonumber\\
P_{\mu e}&=4|V_{\mu4}|^2|V_{e4}|^2\sin^2\Delta_{41}.
\label{eq:sbl_prob}
\end{align}
These expressions require one resolved oscillation frequency, but no small-mixing approximation. They also hold for antineutrinos: conjugating $V$ leaves their coefficients unchanged. Intrinsic CP asymmetry cannot arise from this single-frequency vacuum limit.

The effective channel amplitudes are defined by
\begin{align}
\sin^2 2\theta_{ee}&\equiv4|V_{e4}|^2(1-|V_{e4}|^2),\nonumber\\
\sin^2 2\theta_{\mu\mu}&\equiv4|V_{\mu4}|^2(1-|V_{\mu4}|^2),\nonumber\\
\sin^2 2\theta_{\mu e}&\equiv4|V_{\mu4}|^2|V_{e4}|^2.
\label{eq:sin_mue}
\end{align}
The effective angles $\theta_{ee}$, $\theta_{\mu\mu}$, and $\theta_{\mu e}$ label disappearance and appearance amplitudes; they are not independent rotation angles in $V$. Set $x\equiv|V_{e4}|^2$ and $y\equiv|V_{\mu4}|^2$, the electron and muon fractions of $\nu_4$. For $x,y<1$,
\begin{equation}
\sin^2 2\theta_{\mu e}
=\frac{\sin^2 2\theta_{ee}\,\sin^2 2\theta_{\mu\mu}}{4(1-x)(1-y)}
\simeq\frac14\sin^2 2\theta_{ee}\,\sin^2 2\theta_{\mu\mu}
\quad(x,y\ll1).
\label{eq:appdis}
\end{equation}
The last step adds the small-mixing assumption. On this branch, two small disappearance amplitudes imply an even smaller appearance amplitude. A common fit must respect this relation. It explains the appearance--disappearance tension; its statistical strength still requires the measured spectra and correlated uncertainties \cite{Dentler:2018sju}.

In laboratory units,
\begin{equation}
\Delta_{41}=1.267\,
\frac{\Delta m^2_{41}}{\mathrm{eV}^2}
\frac{L/\mathrm{km}}{E_\nu/\mathrm{GeV}}.
\label{eq:phase_units}
\end{equation}
The numerical coefficient converts the natural-unit phase to the displayed units. An eV$^2$ splitting produces order-one phases over meters at MeV energies or kilometers at GeV energies. Event predictions must average over source size, neutrino energy, baseline, and detector response. Complete phase averaging gives $\langle\sin^2\Delta_{41}\rangle=1/2$, where brackets denote that average. The spectral oscillations disappear, but the correlated rate shifts remain.

Conventional gallium fits favor substantial electron--sterile mixing. Neutrino-4 also reports a signal at several eV$^2$ \cite{Serebrov:2020kmd}, but reactor spectral comparisons exclude much of the relevant region \cite{NEOS2017,DANSS2025,STEREO2023,PROSPECT2025}. KATRIN's 259-day tritium-spectrum analysis tests the electron admixture through decay kinematics within its accessible mass range \cite{KATRIN:2025lph}. MicroBooNE combines appearance and disappearance information. MINOS/MINOS+, NOvA, T2K, and IceCube constrain the muon admixture. Together, these tests strongly disfavor a common minimal $3+1$ explanation \cite{Acero2024}.
\label{sec:sterile_constraints}

Cosmology tests the abundance and mass density of additional states. After electron--positron annihilation, the relativistic energy density is conventionally written as
\begin{equation}
\rho_{\rm rad}=\rho_\gamma\left[1+\frac78
\left(\frac4{11}\right)^{4/3}N_{\rm eff}\right].
\label{eq:neff}
\end{equation}
Here $\rho_{\rm rad}$ is the total radiation energy density, $\rho_\gamma$ is the photon energy density, and $N_{\rm eff}$ measures the remaining relativistic density in units of a standard neutrino species. The factors $7/8$ and $(4/11)^{4/3}$ encode Fermi statistics and the reference neutrino-to-photon temperature ratio. The standard prediction is close to three. In a conventional radiation-dominated history with small lepton asymmetry, anomaly-scale mixing can populate a sterile state before active-neutrino decoupling \cite{Thermalization2012}. Its extra contribution $\Delta N_{\rm eff}\equiv N_{\rm eff}-N_{\rm eff}^{\rm std}$ can approach one, where $N_{\rm eff}^{\rm std}$ is the standard prediction. Once nonrelativistic, the population also affects structure growth through its mass and free streaming.

For a thermal sterile distribution with zero chemical potential and otherwise standard active distributions,
\begin{equation}
\Delta N_{\rm eff}=\left(\frac{T_s}{T_\nu}\right)^4,
\qquad
m_s^{\rm eff}=\left(\frac{T_s}{T_\nu}\right)^3m_4
=(\Delta N_{\rm eff})^{3/4}m_4.
\label{eq:meff}
\end{equation}
The temperatures $T_s$ and $T_\nu$ describe the sterile and standard neutrino distributions. The effective mass $m_s^{\rm eff}$ parametrizes the sterile nonrelativistic mass density relative to a standard thermal neutrino number density. It differs from the physical mass $m_4$ unless the populations match. The conversion changes for a nonthermal distribution.

Cosmic microwave background (CMB) and baryon acoustic oscillation data strongly disfavor a fully thermalized eV-mass state in standard cosmological extensions \cite{Planck2018,SterileCosmo2026}. The 2026 reassessment using the Dark Energy Spectroscopic Instrument's second data release separates full and partial thermalization and tests different dark-energy models \cite{SterileCosmo2026}. These constraints cannot be replaced by a mass-sum limit derived for three standard active neutrinos.

Low reheating temperatures, lepton asymmetries, or new interactions can suppress sterile production \cite{Abazajian2012,Acero2024}. Each proposal must also satisfy nucleosynthesis and late-time constraints. Suppression at one epoch does not rule out later repopulation. Cosmological limits therefore depend on a specified production history as well as laboratory mixing.

\subsection{Non-standard neutrino interactions}
\label{sec:nsi}
NSI can modify production, detection, or coherent propagation through matter \cite{Miranda2015,Dev2019,Ohlsson2013,FarzanTortola2018}. A microscopic model may relate these effects, but each enters different observables. We consider vector currents containing left-handed neutrinos. Scalar, pseudoscalar, and tensor operators require additional process-dependent terms.

A conventional low-energy NC Lagrangian density is
\begin{equation}
\mathcal L_{\rm NSI}^{\rm NC}=-2\sqrt2G_F
\sum_{f,P,\alpha,\beta}\epsilon_{\alpha\beta}^{fP}
(\bar\nu_\alpha\gamma^\lambda P_L\nu_\beta)
(\bar f\gamma_\lambda P f),
\qquad \epsilon_{\alpha\beta}^{fP}=(\epsilon_{\beta\alpha}^{fP})^*.
\label{eq:lagr_nsi}
\end{equation}
The Fermi constant $G_F$ has units of inverse energy squared. The dimensionless coefficients $\epsilon_{\alpha\beta}^{fP}$ specify the interaction strength relative to $G_F$. Here $\alpha,\beta$ label active flavors and $f=e,u,d$ denotes an electron, up-quark, or down-quark field. The projectors are $P_L=(1-\gamma_5)/2$ and $P_R=(1+\gamma_5)/2$, with $P$ summed over both. The matrices $\gamma^\lambda$ are Dirac matrices, $\gamma_5$ is the chirality matrix, and the repeated Lorentz index $\lambda$ is contracted. A bar on a fermion field denotes its Dirac adjoint. The stated relation between coefficients makes the flavor sum Hermitian.

For semileptonic CC interactions, an illustrative term is
\begin{equation}
\mathcal L_{\rm NSI}^{\rm CC}=-2\sqrt2G_F V_{ud}^{\rm CKM}
\sum_{P,\alpha,\beta}\epsilon_{\alpha\beta}^{udP}
(\bar\ell_\alpha\gamma^\lambda P_L\nu_\beta)
(\bar u\gamma_\lambda P d)+\mathrm{h.c.}
\label{eq:cc_nsi}
\end{equation}
The field $\ell_\alpha$ is the charged lepton of flavor $\alpha$. The coefficients $\epsilon_{\alpha\beta}^{udP}$ are dimensionless CC couplings, and h.c. denotes the Hermitian-conjugate term. The Cabibbo--Kobayashi--Maskawa matrix element $V_{ud}^{\rm CKM}$ mixes up and down quarks; it is distinct from the neutrino matrix $V$. Muon-decay sources require purely leptonic operators. Matching these low-energy terms to an electroweak gauge-invariant theory can relate them to charged-lepton observables and precision weak decays \cite{Biggio:2009nt,Falkowski:2021bkq}.

For three active flavors in unpolarized matter, the flavor-basis Hamiltonian is
\begin{align}
H_\nu&=\frac{1}{2E_\nu}U\operatorname{diag}(m_1^2,m_2^2,m_3^2)U^\dagger+V_{\rm mat},\nonumber\\
V_{\rm mat}&=\sqrt2G_FN_e\left[\operatorname{diag}(1,0,0)+\epsilon^m\right],\nonumber\\
\epsilon^m_{\alpha\beta}&=\sum_{f=e,u,d}\frac{N_f}{N_e}
\left(\epsilon^{fL}_{\alpha\beta}+\epsilon^{fR}_{\alpha\beta}\right).
\label{eq:matter_nsi}
\end{align}
The operator $\operatorname{diag}$ forms a diagonal matrix. The quantities $N_f$ and $N_e$ are the number densities of matter fermion $f$ and electrons. The Hermitian matrix $V_{\rm mat}$ is the coherent forward-scattering potential, and $\epsilon^m$ is its dimensionless NSI contribution. The superscript $m$ denotes matter; $L,R$ on the microscopic coefficients denote chirality. The common standard NC potential has been subtracted because only diagonal differences affect active-flavor oscillations. For antineutrinos, $U\to U^*$ and $V_{\rm mat}\to-V_{\rm mat}^*$. A varying density requires path-ordered evolution.

Matter NSI can shift LBL and atmospheric probabilities and correlate with $\delta_{\rm CP}$, $\theta_{23}$, and the ordering. For terrestrial SBL paths, allowed ordinary-matter couplings generally accumulate too little phase to explain the appearance excesses. Production and detection NSI require a separate treatment.

In a factorized pure-state description, define the dimensionless source and detector matrices
\begin{equation}
S_{\alpha\gamma}=\delta_{\alpha\gamma}+\epsilon^s_{\alpha\gamma},
\qquad
D_{\gamma\beta}=\delta_{\gamma\beta}+\epsilon^d_{\gamma\beta}.
\label{eq:sd_matrices}
\end{equation}
The index $\gamma$ runs over active flavors. The coefficient $\epsilon^s_{\alpha\gamma}$ adds a flavor-$\gamma$ component to a source labeled by charged-lepton flavor $\alpha$. The coefficient $\epsilon^d_{\gamma\beta}$ changes the amplitude for detecting flavor $\gamma$ through charged lepton $\beta$. The superscripts $s,d$ denote source and detector, not a sterile flavor. With the detector coefficient placed in the bra \cite{Kopp2008,OhlssonZhang2009}, the normalized states are
\begin{align}
|\nu_\alpha^s\rangle&=\frac{1}{N_\alpha^s}\sum_\gamma S_{\alpha\gamma}|\nu_\gamma\rangle,
&\langle\nu_\beta^d|&=\frac{1}{N_\beta^d}\sum_\gamma D_{\gamma\beta}\langle\nu_\gamma|,\label{eq:nsi_states}\\
(N_\alpha^s)^2&=(SS^\dagger)_{\alpha\alpha},
&(N_\beta^d)^2&=(D^\dagger D)_{\beta\beta}.
\label{eq:nsi_norms}
\end{align}
The normalization factors $N_\alpha^s$ and $N_\beta^d$ are positive real numbers. They are unrelated to the matter densities $N_f$. The effective coefficients depend on the production and detection matrix elements. They are generally different from the operator coefficients in Eqs.~\eqref{eq:lagr_nsi} and~\eqref{eq:cc_nsi}. Entanglement with unobserved particles or nonfactorizing amplitudes can require a density-matrix treatment instead of these pure states.

Let $\mathcal U(L)$ be the unitary coherent evolution operator over distance $L$. Define the normalized transition overlap $P_{\alpha\beta}^{sd}$ and the dimensionless event-rate kernel $K_{\alpha\beta}$ by
\begin{equation}
P_{\alpha\beta}^{sd}=|\langle\nu_\beta^d|\mathcal U(L)|\nu_\alpha^s\rangle|^2,
\qquad
K_{\alpha\beta}=(N_\alpha^s)^2(N_\beta^d)^2P_{\alpha\beta}^{sd}.
\label{eq:nsi_kernel}
\end{equation}
The overlap is bounded by one. The kernel also contains production and detection normalization changes, so it need not be. Detector states can be nonorthogonal; their squared overlaps do not form a normalized set of exclusive outcomes. Physical rates follow from the interaction amplitudes.

At zero propagation distance, for $\alpha\ne\beta$ and small source/detector coefficients,
\begin{align}
P_{\alpha\beta}^{sd}(0)&=\frac{|(SD)_{\alpha\beta}|^2}{(N_\alpha^s)^2(N_\beta^d)^2}
=|\epsilon^s_{\alpha\beta}+\epsilon^d_{\alpha\beta}|^2+\mathcal O(\epsilon^3),\label{eq:zero_distance}\\
|\epsilon^s_{\alpha\beta}+\epsilon^d_{\alpha\beta}|^2
&=|\epsilon^s_{\alpha\beta}|^2+|\epsilon^d_{\alpha\beta}|^2
+2|\epsilon^s_{\alpha\beta}||\epsilon^d_{\alpha\beta}|
\cos(\phi^s_{\alpha\beta}-\phi^d_{\alpha\beta}).
\label{eq:zero_phase}
\end{align}
Here $\epsilon$ denotes the common small scale of the coefficients, and $\mathcal O(\epsilon^3)$ collects cubic and higher orders. Their phases are defined by $\epsilon^{s,d}_{\alpha\beta}=|\epsilon^{s,d}_{\alpha\beta}|e^{i\phi^{s,d}_{\alpha\beta}}$. The phase difference follows from the detector-bra convention. The nonzero overlap arises at production or detection; it is not accumulated along the path.

Diagonal coefficients illustrate the importance of normalization. If only $\epsilon^s_{\mu\mu}$ and $\epsilon^d_{\mu\mu}$ are nonzero, then $P^{sd}_{\mu\mu}=P^{\rm std}_{\mu\mu}$, where $P^{\rm std}_{\mu\mu}$ is the standard muon-neutrino survival probability. The rate kernel is instead
\begin{align}
K_{\mu\mu}&=|1+\epsilon^s_{\mu\mu}|^2|1+\epsilon^d_{\mu\mu}|^2P^{\rm std}_{\mu\mu}\nonumber\\
&\simeq\left[1+2\operatorname{Re}(\epsilon^s_{\mu\mu}+\epsilon^d_{\mu\mu})\right]
\left[1-\sin^22\theta_{23}\sin^2\!\left(\frac{\Delta m^2_{31}L}{4E_\nu}\right)\right].
\label{eq:nsi_disappearance}
\end{align}
The first line is exact within the stated source/detector model. The second keeps only first-order NSI terms and uses the two-flavor vacuum limit: $\theta_{13}=0$ and negligible solar phase. The symbol $\operatorname{Re}$ denotes the real part. Diagonal NSI change the rate even when the normalized survival probability is unchanged. Flavor-changing coefficients add interference and spectral effects.

For a reconstructed-energy bin $b$, a factorized prediction takes the form
\begin{equation}
N_b=\mathcal E\sum_{\alpha,\beta}\int dE_\nu\,dL\,
\Phi_\alpha^0(E_\nu,L)\,\sigma_\beta^0(E_\nu)\,
K_{\alpha\beta}(E_\nu,L)\,R_b^\beta(E_\nu,L)+B_b.
\label{eq:eventrate}
\end{equation}
Here $N_b$ is the expected event count and $B_b$ is the expected background in that bin. The exposure $\mathcal E$ is the number of targets times live time. The quantity $\Phi_\alpha^0(E_\nu,L)dL$ is the reference differential incident flux from the indicated baseline interval, including source geometry. The cross section $\sigma_\beta^0$ describes the reference detection process. The dimensionless response $R_b^\beta$ includes selection efficiency and migration into bin $b$. Superscript $0$ denotes the chosen Standard Model reference, with NSI normalization changes contained in $K_{\alpha\beta}$. If factorization fails, full differential production and detection amplitudes are required.

An energy-independent transition overlap does not produce a flat reconstructed spectrum. Flux, cross section, and response all depend on energy. MiniBooNE and MicroBooNE have different responses and backgrounds even in the same beam. An energy-independent correction shared by identical source and detector processes can instead cancel approximately in ratios between baselines.

Source and detector NSI are constrained by weak-decay rates, lepton universality, quark-mixing consistency, and near-detector flavor measurements \cite{Biggio:2009nt,Falkowski:2021bkq}. Each bound applies to a specified operator, flavor combination, and interference pattern. Daya Bay's full-data CC analysis illustrates this dependence \cite{DayaBayCCNSI2024}. In its correlated source/detector parametrization, $\epsilon_{ee}$ denotes the common diagonal electron-flavor coefficient. With this coefficient alone nonzero and its phase fixed to zero, the result is $|\epsilon_{ee}|<0.0148$ at 90\% confidence for one degree of freedom. Freeing the phase permits much larger degenerate regions. Alternative source and effective-operator parametrizations give different limits, including selected bounds of order $10^{-1}$.
\label{sec:nsi_constraints}

NOMAD's null appearance result constrains operators that act in its high-energy production and detection channels \cite{NOMAD:2003mqg}. LSND instead uses muon decay and IBD. Explaining its mean probability solely through Eq.~\eqref{eq:zero_distance} requires an effective source-plus-detector amplitude near $\sqrt{0.00264}\simeq0.05$. KARMEN and precision constraints on the relevant flavor-violating operators then provide direct tests \cite{Bergmann:1998ft}. Each experiment still requires a prediction folded with its own flux, cross section, and response.

For matter NSI, NOvA varied $\epsilon^m_{e\mu}$ or $\epsilon^m_{e\tau}$ separately, including the phase of the selected coefficient \cite{NOvA2024NSI}. The 90\% regions near the standard solution give approximately $|\epsilon^m_{e\mu}|\lesssim0.3$ and $|\epsilon^m_{e\tau}|\lesssim0.4$. A further degenerate region appears near $|\epsilon^m_{e\tau}|\simeq1.8$. These are single-coupling results. Earlier T2K/NOvA phenomenological fits improved with new phases and couplings of order $0.1$--$0.2$ \cite{Denton:2020uda,Chatterjee:2020kkm,Chatterjee2024}. The later joint standard analysis does not require such a contribution.

IceCube DeepCore tests generalized matter potentials at GeV energies \cite{IceCubeNSI2021}. A TeV analysis gives $-0.0041\leq\epsilon^m_{\mu\tau}\leq0.0031$ at 90\% confidence when that effective Earth-matter coefficient is real \cite{IceCubeMuTauNSI2022}. Extending this interval to complex simultaneous couplings, or to individual electron and quark coefficients, requires a new fit.

Oscillations admit a generalized ordering degeneracy under $H_\nu\to-H_\nu^*$, up to a flavor-independent term. In NSI models it can map the solar angle into the opposite octant, $\theta_{12}>\pi/4$. This is the large-mixing-angle dark solution, or LMA-Dark \cite{Esteban:2018ppq}. Coherent elastic neutrino--nucleus scattering (CE$\nu$NS) probes other coupling combinations and reduces the degeneracy \cite{Coloma2023,ColomaNSI2020}. The result depends on mediator mass, target composition, and cancellations among matter couplings. A 2025 solar CE$\nu$NS study found surviving examples with mediators coupled to muon/tau neutrinos and quarks \cite{LMADark2025}. For light mediators, translating a forward-scattering potential into a finite-momentum scattering limit requires the propagator.

A common NSI explanation must therefore connect production, propagation, detection, and scattering within one operator model. Proposed DUNE near-detector tests can improve this comparison \cite{Giarnetti2021}; their sensitivities remain projections.

\subsection{Decoherence and open-system dynamics}
\label{sec:decoherence}
Unobserved degrees of freedom can become correlated with a neutrino and reduce its observable coherence. The neutrino is then described by a density matrix $\rho$, a positive operator normalized by $\operatorname{Tr}\rho=1$, where $\operatorname{Tr}$ denotes the trace. The combined neutrino--environment system can remain unitary while the reduced neutrino state evolves non-unitarily \cite{BenattiFloreanini2000,LisiMarroneMontanino2000}.

A common description is the Gorini--Kossakowski--Sudarshan--Lindblad equation \cite{GKS1976,Lindblad1976}:
\begin{equation}
\frac{d\rho}{dt}=-i[H,\rho]+\mathcal D[\rho],
\qquad
\mathcal D[\rho]=\sum_a\left(L_a\rho L_a^\dagger
-\frac12\{L_a^\dagger L_a,\rho\}\right).
\label{eq:lindblad}
\end{equation}
Here $t$ is propagation time, $H$ is the coherent Hamiltonian, and $\mathcal D$ is the dissipator. The index $a$ labels Lindblad operators $L_a$, which describe incoherent processes and have units of energy$^{1/2}$. Square and curly brackets denote the commutator and anticommutator. The symbol $L_a$ denotes an operator, whereas $L$ remains the propagation distance. This equation preserves the trace and generates completely positive Markovian evolution. Complete positivity preserves valid probabilities even when the neutrino is correlated with an auxiliary system.

Constant coefficients describe a time-homogeneous semigroup. This assumes negligible environmental memory on the resolved propagation scales. Weak-coupling derivations usually also require short environmental correlation times and coarse graining. Invisible loss or scattering out of an accepted momentum bin requires a larger state space or a transport equation. Trace preservation applies only to the state space actually included.

Consider vacuum propagation with three active flavors. Choose Hermitian Lindblad operators diagonal in the mass basis:
\begin{equation}
H_{ij}=\delta_{ij}\frac{m_i^2}{2E_\nu},\qquad
(L_a)_{ij}=\delta_{ij}\ell_{ai},\qquad \ell_{ai}\in\mathbb R.
\label{eq:dephase_ops}
\end{equation}
The indices $i,j=1,2,3$ label mass states, $\delta_{ij}$ is the Kronecker delta, and $\ell_{ai}$ is the real eigenvalue of $L_a$ on state $i$. The operators commute with $H$, so the diagonal populations $\rho_{ii}$ stay constant. The off-diagonal elements evolve as
\begin{equation}
\rho_{ij}(L)=\rho_{ij}(0)
\exp\!\left[-i\frac{\Delta m^2_{ij}L}{2E_\nu}-\Gamma_{ij}(E_\nu)L\right],
\qquad
\Gamma_{ij}=\frac12\sum_a(\ell_{ai}-\ell_{aj})^2,
\quad i\ne j.
\label{eq:rho_dephase}
\end{equation}
We use the ultrarelativistic relation $t\simeq L$. The nonnegative rate $\Gamma_{ij}=\Gamma_{ji}$ damps coherence between states $i$ and $j$ and has units of energy, or inverse length. The square-difference form guarantees complete positivity. The quantities $\sqrt{\Gamma_{ij}}$ are proportional to distances between vectors of Lindblad eigenvalues and obey triangle inequalities. Thus the three rates are not arbitrary. For example, damping only the $12$ coherence while preserving $13$ and $23$ is inconsistent with this construction.

For an initial flavor state $\rho(0)=|\nu_\alpha\rangle\langle\nu_\alpha|$, projection onto flavor $\beta$ gives
\begin{align}
P_{\alpha\beta}(L,E_\nu)
&=\sum_i|U_{\alpha i}|^2|U_{\beta i}|^2
+2\sum_{i<j}e^{-\Gamma_{ij}(E_\nu)L}
\operatorname{Re}\!\left[J_{\alpha\beta}^{ij}
 e^{-i\Delta m^2_{ij}L/(2E_\nu)}\right],\label{eq:deco_prob}\\
J_{\alpha\beta}^{ij}&\equiv U_{\alpha i}^*U_{\alpha j}U_{\beta i}U_{\beta j}^*.
\label{eq:deco_J}
\end{align}
The quantity $J_{\alpha\beta}^{ij}$ is the mixing-matrix product multiplying the $i,j$ interference term. Only interference is damped. Vanishing rates recover standard oscillations; finite rates give $P_{\alpha\beta}(0)=\delta_{\alpha\beta}$ and $\sum_\beta P_{\alpha\beta}=1$. Complete dephasing gives the limit $\sum_i|U_{\alpha i}|^2|U_{\beta i}|^2$. This is generally different from equal flavor probabilities of $1/3$. Equalization requires additional population-changing dynamics.

For two flavors, the appearance probability reduces to
\begin{equation}
P_{\alpha\beta}=\frac{\sin^22\theta}{2}
\left[1-e^{-\Gamma L}\cos\!\left(\frac{\Delta m^2 L}{2E_\nu}\right)\right],
\qquad \alpha\ne\beta.
\label{eq:deco_two}
\end{equation}
Here $\theta$ is the two-flavor mixing angle, $\Delta m^2$ is the mass-squared splitting of the two states, and $\Gamma$ is their coherence-damping rate. If the coherent phase is negligible and $\Gamma L\ll1$, the leading term is $P_{\alpha\beta}\simeq\tfrac12\sin^22\theta\,\Gamma L$. The omitted coherent contribution is quadratic in the oscillation phase. Dephasing can therefore generate appearance at finite baseline even when the standard phase is small. The effect vanishes at zero distance. Depending on the coherent phase, damping can either raise or lower a probability.

More general Lindblad operators also change populations or mix density-matrix components \cite{BarenboimGago2024}. Equation~\eqref{eq:deco_prob} does not cover those cases. Dephasing in the instantaneous matter basis is a further model choice. In a varying medium, both the Hamiltonian and dissipator must be evolved consistently \cite{CarpioMassoniGago2018}. Microscopic scattering can also change neutrino momentum. A 2026 field-theoretic treatment connects such dissipative terms to scattering cross sections \cite{Stankevich2026}; that mapping is needed before identifying a fitted damping rate with a fundamental coupling.

The energy dependence is often parametrized as
\begin{equation}
\Gamma_{ij}(E_\nu)=\gamma_{ij}\left(\frac{E_\nu}{E_0}\right)^n,
\qquad E_0=1~\mathrm{GeV},\qquad \gamma_{ij}\equiv\Gamma_{ij}(E_0).
\label{eq:gamma_power}
\end{equation}
The fixed reference energy is $E_0$, the dimensionless exponent is $n$, and $\gamma_{ij}$ is the rate at that reference energy. Each $\gamma_{ij}$ has energy units, regardless of $n$. Since $1~\mathrm{km}=5.0677\times10^{18}~\mathrm{GeV}^{-1}$, the damping exponent $\gamma_{ij}(E_\nu/E_0)^nL$ is dimensionless. Positive $n$ enhances high-energy sensitivity; negative $n$ enhances low-energy sensitivity. For $n=-1$, damping scales as $L/E_\nu$. This scaling alone does not establish Lorentz invariance of an underlying environment. A common energy factor preserves a valid rate pattern; different energy laws must satisfy positivity at every energy.

``Soft decoherence'' models fitted LSND by choosing rapidly varying energy dependence and selected damped coherences \cite{Farzan2008,Farzan2015}. They show that three-active-state open dynamics can generate SBL appearance. Their fits predate the full present dataset. Testing them now requires the specified rate pattern, reactor constraints, and neutrino/antineutrino prescription across KARMEN, MiniBooNE, JSNS$^2$, and MicroBooNE.

Farzan and Schwetz also proposed a gallium model with a characteristic damping length $1/\Gamma$ of order 2~m at source energies \cite{FarzanGallium2023}. A steep decrease, exemplified by $\Gamma\propto E_\nu^{-12}$, avoids higher-energy bounds. Changing source energy or path-length coverage directly tests this construction. Its required damping also depends on the capture cross section discussed in Section~\ref{sec:gallium}.

At long baselines, neglected damping can bias inferred mixing parameters \cite{CarpioMassoniGago2019}. A successful interpretation must fit appearance and disappearance in both beam polarities with one dissipator. It must also survive reactor and atmospheric constraints. The joint T2K--NOvA result does not establish a need for damping.

Existing spectral data already set strong model-specific limits. De Romeri et al. analyzed RENO, Daya Bay, KamLAND, T2K, NOvA, and MINOS/MINOS+ \cite{DeRomeri2023}. In their model A, the three rates have a common normalization $\gamma\equiv\gamma_{21}=\gamma_{31}=\gamma_{32}$ at $E_0=1$~GeV. The strongest reported 90\% limits are $\gamma<7.8\times10^{-27}$~GeV for $n=-2$ from KamLAND and $\gamma<5.1\times10^{-24}$~GeV for $n=0$ from MINOS/MINOS+. These bounds describe different energy laws. Comparisons also require compatible rate patterns and matter-basis prescriptions.

KM3NeT/ORCA's six-unit analysis found no significant deviation \cite{ORCA2025}. It uses dephasing in the effective matter basis and imposes $\gamma_{32}=(\sqrt{\gamma_{31}}-\sqrt{\gamma_{21}})^2$, an additional restriction beyond general complete positivity. At $E_0=1$~GeV, the conservative 90\% limits over the two orderings are $\gamma_{21}<1.9\times10^{-22}$~GeV and $\gamma_{31}<2.7\times10^{-22}$~GeV for $n=-1$. For $n=-2$, they are $4.6\times10^{-21}$ and $8.4\times10^{-21}$~GeV, respectively. Each one-dimensional bound profiles over the other free damping parameter.

IceCube's 2024 analysis of 0.5--10~TeV atmospheric neutrinos found no signal in its phase-perturbation and state-selection scenarios \cite{IceCubeDeco2024}. These high energies strongly test positive powers of energy. A later 10.7-year study appeared in a 2025 sensitivity contribution and 2026 conference presentations \cite{IceCubeDecoProspect2025,IceCubeDecoTalk2026}. Those projections and preliminary reports are distinct from the published 2024 result.

Solar limits depend on matter evolution. In the two-flavor solar/KamLAND model of Fogli et al., $\gamma_0$ is the damping normalization at 1~GeV in the vacuum energy law; their solar prescription also includes the density profile. They obtain $\gamma_0<7.8\times10^{-27}$~GeV at 95\% confidence for $n=-1$ \cite{Fogli2007}. This is not a universal limit on an arbitrary vacuum-dephasing rate. At astrophysical distances, ordinary phase averaging may already remove all interference. Additional pure mass-basis dephasing then leaves the same flavor limit, while relaxation or loss can still change flavor ratios and yields. The cited SN1987A analysis includes neutrino loss and depends on supernova emission and propagation assumptions \cite{SNDeco2025}.

Loss of oscillation contrast can also arise without a new environmental interaction. In a simple Gaussian wave-packet model, separated mass components produce a factor
\begin{equation}
\exp\!\left[-\left(\frac{L}{L^{\rm coh}_{ij}}\right)^2\right],
\qquad
L^{\rm coh}_{ij}=\frac{4\sqrt2 E_\nu^2\sigma_x}{|\Delta m^2_{ij}|}.
\label{eq:wavepacket}
\end{equation}
The coherence length $L^{\rm coh}_{ij}$ is the distance over which packets $i,j$ separate appreciably. The effective spatial width $\sigma_x$ uses the convention in which a Gaussian packet amplitude has an exponent equal to minus the squared displacement divided by $4\sigma_x^2$. Production, detection, and localization determine the physical width and any additional suppression factors \cite{Akhmedov2022,Krueger2023}. The Gaussian baseline dependence differs from constant-rate Lindblad damping.

Finite energy resolution, source size, and baseline coverage instead average a coherent probability over unresolved events. Adding another damping factor for the same averaging would double count it. Quantum-field-theoretic changes to mixing or curved-background phases likewise require a specified reduction of the state before they can be interpreted as environmental decoherence \cite{SeraoCurved2023,SeraoTorsion2024,Capolupo:2023kvy,Capolupo:2020wlx}.

\subsection{Complementary interpretations}
\label{sec:complementary}
Heavy neutral leptons that cannot be produced can leave an effectively non-unitary light-neutrino mixing matrix. Weak rates, flavor-state overlaps, and matter effects then change together \cite{Blennow2017}. Precision weak observables and near-detector normalization constrain this possibility.

Neutrino decay changes the populations during propagation. For an ultrarelativistic unstable mass eigenstate $\nu_i$, the vacuum survival amplitude contains $\exp[-m_iL/(2E_\nu\tau_i)]$, where $\tau_i$ is its rest-frame lifetime. The survival probability contains twice that exponent. Detectable daughters redistribute energy; invisible decay removes events from the observed sector. IceCube's 2026 analysis strongly constrains the specific invisibly decaying eV-state regions motivated by SBL fits \cite{IceCubeDecay2026}.

Dark-sector upscattering and decay can generate photons or electron--positron ($e^+e^-$) showers \cite{Bertuzzo2018}. Relevant tests include invariant masses, opening angles, hadronic recoil, displaced vertices, and target dependence. Predictions must be simulated through each detector's event selection. An oscillation exclusion contour alone does not test these processes.

\section{Discriminating Signatures and Experimental Outlook}
\label{sec:outlook}
\subsection{Comparison of mechanisms}
No common origin of the reviewed anomalies has been established. LSND and MiniBooNE remain distinct unexplained excesses. Reactor rate deficits depend on flux models, and gallium deficits depend on capture calculations. The joint T2K--NOvA analysis is compatible with three-flavor oscillations. Table~\ref{tab:comparison} connects each mechanism to its signatures and independent tests. It summarizes existing evidence rather than ranking models through a new fit.

\begin{table}[H]
\centering
\caption{Mechanisms, signatures, and independent tests. SBL and LBL denote short and long baselines; NSI denotes non-standard neutrino interactions; NC denotes neutral current; CE$\nu$NS denotes coherent elastic neutrino--nucleus scattering. The baseline is $L$, the neutrino energy is $E_\nu$, and $\Gamma_{ij}$ is the damping rate of coherence between states $i,j$.}
\label{tab:comparison}
\footnotesize
\renewcommand{\arraystretch}{1.16}
\setlength{\tabcolsep}{3pt}
\begin{tabularx}{\dimexpr\linewidth-0.4cm\relax}{@{}>{\raggedright\arraybackslash}p{0.19\linewidth}YY@{}}
\toprule
\textbf{Mechanism} & \textbf{Physical change and signature} & \textbf{Observations and independent tests}\\
\midrule
Light sterile neutrino ($3+1$) &
\textbf{Physical change:} One extra mass state and oscillation frequency. \textbf{Signature:} Common $L/E_\nu$ frequency; correlated appearance and disappearance; Earth resonances. &
\textbf{Observations:} LSND/MiniBooNE; gallium and reactor deficits under conventional rate predictions. \textbf{Independent tests:} Reactor spectral ratios; muon disappearance; MicroBooNE; KATRIN; cosmology.\\
\addlinespace
Source/\allowbreak detector NSI &
\textbf{Physical change:} Modified production or detection amplitudes. \textbf{Signature:} Flavor admixtures at the source or detector; rate shifts; process-dependent spectra. &
\textbf{Observations:} SBL appearance and normalization changes. \textbf{Independent tests:} Weak decays; universality; near detectors; electroweak matching.\\
\addlinespace
Matter NSI &
\textbf{Physical change:} Additional coherent forward-scattering potential. \textbf{Signature:} Density and beam-polarity dependence; new phases; ordering degeneracies. &
\textbf{Observations:} LBL, solar, and atmospheric parameter inference. \textbf{Independent tests:} Oscillations in different media; CE$\nu$NS; neutrino--electron scattering.\\
\addlinespace
Pure dephasing &
\textbf{Physical change:} Suppressed mass-basis coherence with fixed populations. \textbf{Signature:} Damping $e^{-\Gamma_{ij}(E_\nu)L}$; standard zero-distance limit; mixing-dependent asymptotic probabilities. &
\textbf{Observations:} Specific LSND or gallium models; distorted oscillation contrast. \textbf{Independent tests:} Reactor, accelerator, solar, and atmospheric spectra; positivity of the rate pattern.\\
\addlinespace
Population-changing open dynamics &
\textbf{Physical change:} Relaxation, momentum redistribution, or loss from an observed sector. \textbf{Signature:} Changes in flavor composition, total yield, or energy distribution. &
\textbf{Observations:} Model-specific deficits and propagation distortions. \textbf{Independent tests:} Flavor sums; NC rates; spectra; astrophysical yields; scattering constraints.\\
\bottomrule
\end{tabularx}
\end{table}

\subsection{Discriminating measurements}
Joint baseline and energy measurements test the propagation law. An additional mass splitting generates a phase proportional to $L/E_\nu$. Power-law dephasing instead has an exponent $\gamma_{ij}(E_\nu/E_0)^nL$. Varying energy at fixed $L/E_\nu$ separates these scalings except when $n=-1$; even then, sinusoidal oscillations and exponential damping remain distinct functions. Source/detector NSI can give a nonzero overlap without propagation. Unresolved oscillations can mimic a constant effect, so each experiment must specify its resolved frequency range.

Appearance, disappearance, and both beam polarities must be fitted together. In $3+1$ mixing, Eq.~\eqref{eq:appdis} fixes their amplitude relation. Active-flavor dephasing conserves total active probability. Sterile conversion and invisible loss can instead reduce NC rates. Source or detector NSI can also change the cross sections used in that comparison. Neutrino and antineutrino samples test complex phases, the sign of the matter potential, and any assumed asymmetry in the dissipator.

Event topology tests what produced the excess. Electron/photon separation, visible protons, pion control samples, angular distributions, and displaced or multiple showers constrain different mechanisms. Argon, hydrocarbon, gallium, and electron-scattering targets also probe different nuclear and detector effects. Predictions must be folded with each experiment's response rather than transferred by rescaling total counts.

Near detectors constrain flux and interactions, but can also contain the new-physics effect being tested. Their spectra should be fitted with far-detector data when near-baseline oscillations, source NSI, or rapid damping are possible. Fuel evolution, source-activity measurements, capture calculations, and dedicated interaction data constrain different normalization uncertainties. Response matrices, spectral covariance, and background samples are needed to test more than an integrated rate.

\subsection{Experimental prospects}
The Short-Baseline Neutrino (SBN) program combines SBND and ICARUS at different baselines. SBND has recorded neutrino interactions since 2024; ICARUS has released its first single-detector disappearance result \cite{SBNStatus2026,ICARUS:2026lvh}. Their joint analysis can constrain common beam and argon-interaction uncertainties while testing appearance and disappearance. MicroBooNE's continuing analyses add event-topology information from data collected through 2021 \cite{SBNStatus2026}.

Further JSNS$^2$ exposure will strengthen the direct comparison with LSND. Its second phase includes a farther detector \cite{JSNS2First2026}, allowing a test of baseline dependence. For gallium, sources with different neutrino energies and measurements using neutrino--electron scattering could separate capture uncertainties from flux depletion \cite{ciuffoli2025}. These proposed tests require demonstrated control of their own systematics.

Reactor spectral ratios and isotope-resolved flux measurements remain useful as absolute-rate predictions change. JUNO has already reported oscillation measurements at the literature cutoff \cite{JUNO2026}. Its spectral information can test damping and standard-parameter consistency, provided energy nonlinearity and flux-shape uncertainties are controlled.

The Deep Underground Neutrino Experiment (DUNE) is under construction, with far-detector structural installation progressing in 2026 \cite{DUNEStatus2026}. Its planned wide-band beam, long matter path, and near detector can test matter NSI, source/detector effects, and energy-dependent damping \cite{Giarnetti2021,DeRomeri2023}. Atmospheric data add different densities and a broader energy range. A common fit across these environments can test whether one propagation law explains all channels.

\section{Conclusions}
\label{sec:conclusions}
Three-active-neutrino oscillations remain the leading description of flavor conversion. The unresolved excesses and deficits probe different sources, reactions, and reference predictions. Their interpretation must preserve those differences. The joint T2K--NOvA result also shows that earlier parameter tensions can weaken when experiments are analyzed together.

The minimal $3+1$ model predicts a clear SBL frequency, but ties appearance to electron and muon disappearance. Combined data strongly restrict a common explanation of the anomalies. Reactor and gallium deficits additionally require reliable flux and capture calculations. The recent gallium proposal needs independent nuclear validation before the deficit can be assigned to either propagation or cross-section modeling.

NSI and open-system dynamics broaden the possible mechanisms. Their tests require explicit assumptions: operator and process dependence for NSI; state space, basis, positivity, and energy dependence for decoherence. A normalized overlap, an event rate, and a damped interference term are different quantities. Wave-packet separation and experimental averaging must also be distinguished from environmental dynamics.

No single extension gives an uncontested description of all observations. A successful explanation must reproduce flavor, energy, baseline, and topology measurements with the same parameters while satisfying independent constraints. These correlated tests can determine whether the remaining anomalies reveal new neutrino physics or limitations in the predicted and reconstructed event samples.

\authorcontributions{Author contributions to be confirmed by all authors before submission.}
\funding{This research received no external funding specific to this review.}
\dataavailability{No new experimental data were created or analyzed in this study. The review relies on the published literature cited in the text.}
\acknowledgments{....}
\conflictsofinterest{Conflict-of-interest declarations to be confirmed by all authors before submission.}

\begin{adjustwidth}{-\extralength}{0cm}
\reftitle{References}
\bibliography{bibliography}
\end{adjustwidth}
\end{document}